\documentclass[aps,preprint,epsf,amsmath,amssymb]{revtex4}
\usepackage{latexsym}
\usepackage{amsfonts}
\usepackage{epsfig}

\begin{document}

\title{ Predictions of Dynamic Behavior Under Pressure for Two Scenarios
  to Explain Water Anomalies}

\author{Pradeep Kumar,$^1$ Giancarlo Franzese,$^2$ and
  H. Eugene Stanley$^1$}

\affiliation{$^1$Center for Polymer Studies and Department of
Physics, Boston University -- Boston, MA 02215 USA \\
$^2$Departament de F\'{\i}sica Fonamental -- Universitat de
Barcelona, Diagonal 647, Barcelona 08028,~Spain\\}

\date{last revised 6:30pm 5nov}

%\date{\today} \pacs{05.40.-a}

\begin{abstract}

Using Monte Carlo simulations and mean field calculations for a cell
model of water we find a dynamic crossover in the orientational
correlation time $\tau$ from non-Arrhenius behavior at high temperatures
to Arrhenius behavior at low temperatures.  This dynamic crossover is
independent of whether water at very low temperature is charaterized by
a ``liquid-liquid critical point'' or by the ``singularity free''
scenario.  We relate $\tau$ to fluctuations of hydrogen bond network and
show that the crossover found for $\tau$ for both scenarios is a
consequence of the sharp change in the average number of hydrogen bonds
at the temperature of the specific heat maximum.  We find that the
effect of pressure on the dynamics is strikingly different in the two
scenarios, offering a means to distinguish between them.

\end{abstract}

\maketitle

\begin{figure}
\centerline{\includegraphics[width=8cm]{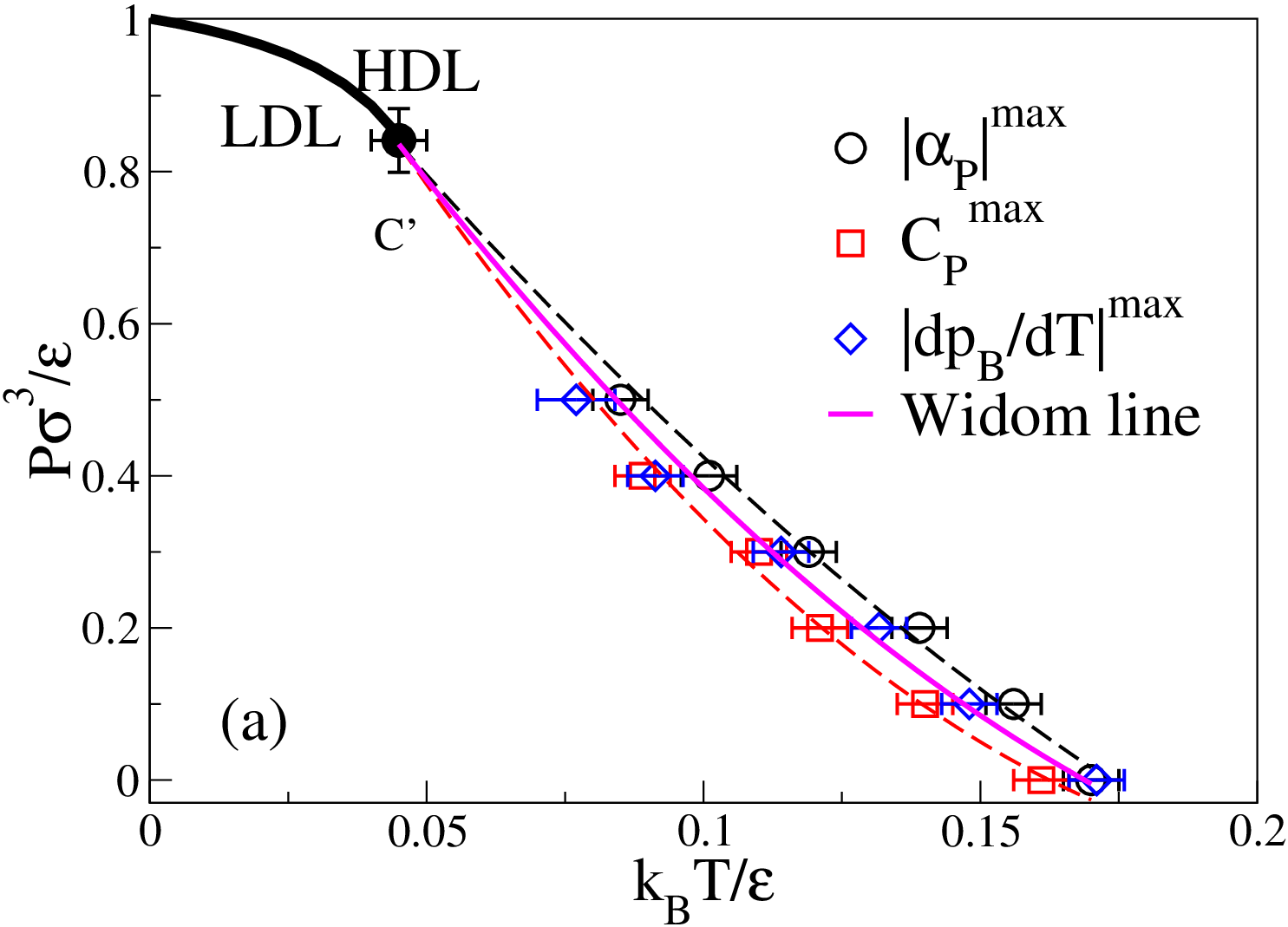}}
%\begin{center}
\includegraphics[width=8cm]{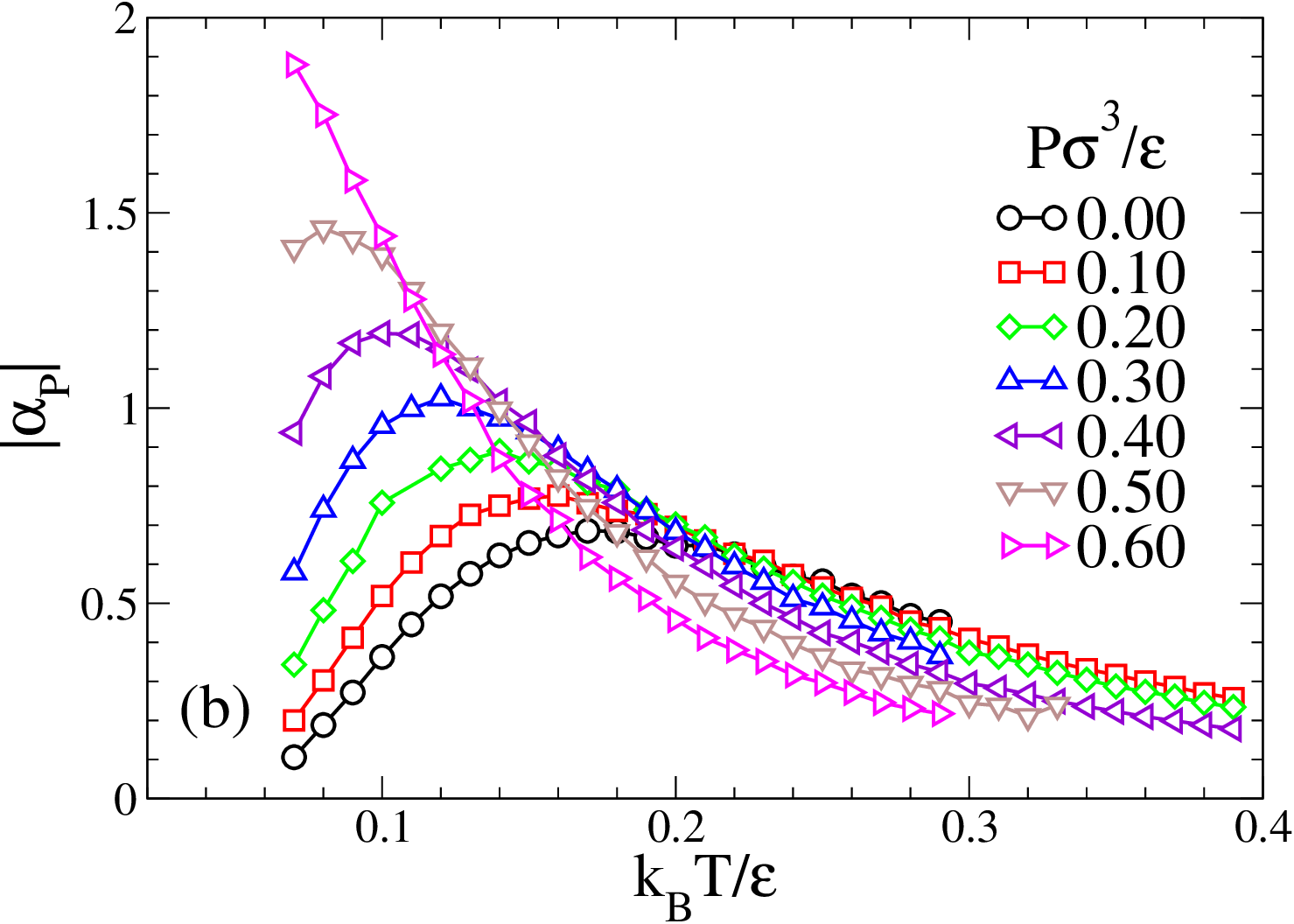}
\includegraphics[width=8cm]{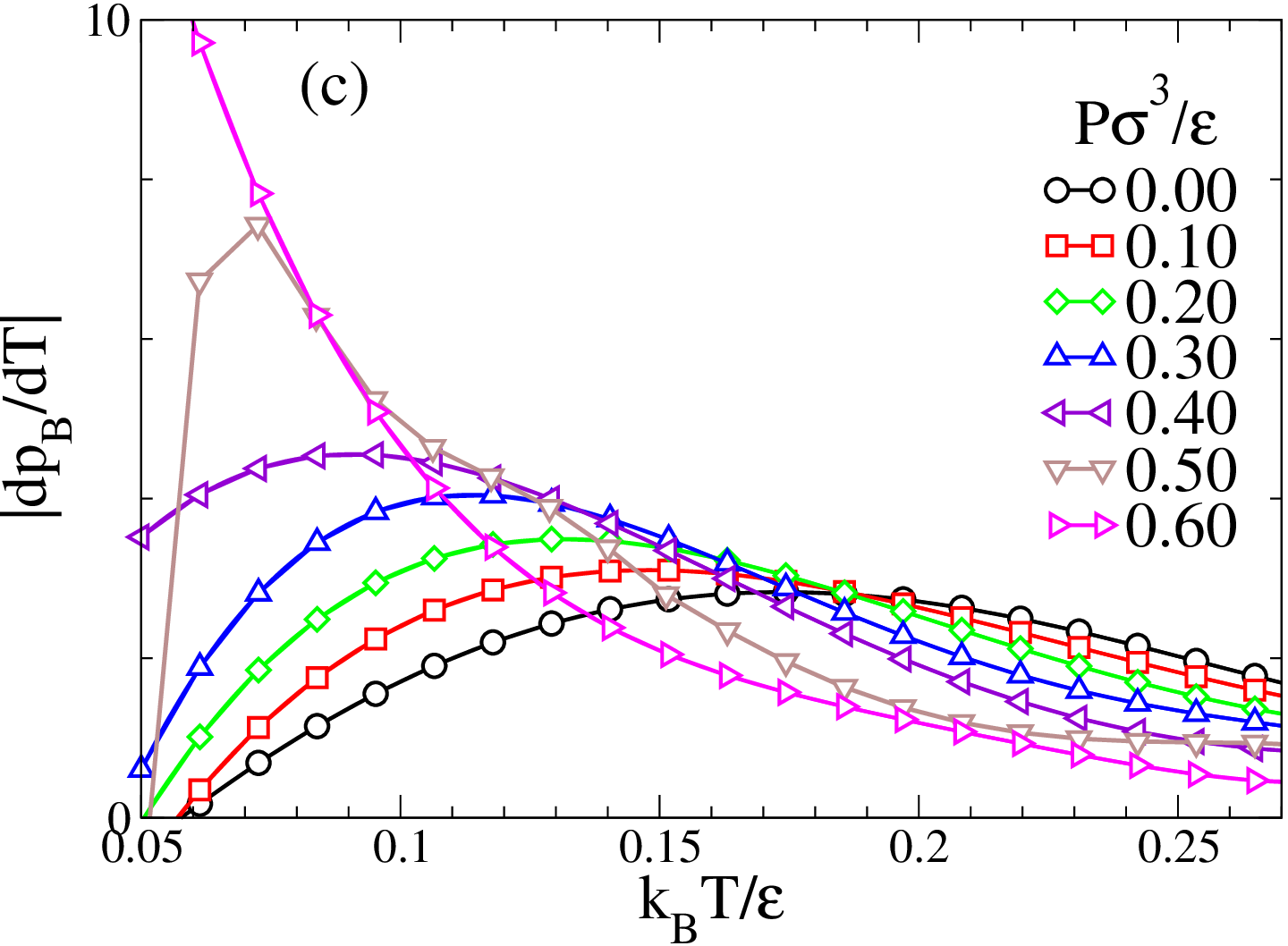}
%\includegraphics[width=4cm]{specific-final-N60.eps}
%\includegraphics[width=4cm]{specific-heat-noIM.eps}
%\includegraphics[width=4cm]{nhb-compar-MF.eps}
%\end{center}
%\singlespacing
\caption{(color online) (a) The relevant part of the liquid phase
  diagram for the LLCP scenario. $C'$ is the LDL-HDL critical point at
  the end of the first-order phase transition line (thick line)
  \cite{fs}, $T_W(P)$ (thin line) is the Widom line, which we take to be
  the average between 
%%%
$T(|\alpha_P|^{\rm
  max})$ ($\bigcirc$)
and 
$T(C_P^{\rm max})$ ($\Box$) 
%%%
%, symbols are
%  the locus of $|\alpha_P|^{\rm max}$ 
%, $C_P^{\rm max}$
%  , and  $|dp_{\rm B}/dT|^{\rm max}$ ($\Diamond$), 
from panel (b), 
%(c) 
and Fig.~\ref{Cp}(a), respectively. 
Upper and lower
  dashed line are quadratic fits of $|\alpha_P|^{\rm max}$ and $C_P^{\rm
  max}$, respectively, merging at the $C'$. 
%%%
$T(|\alpha_P|^{\rm max})$ and $T(C_P^{\rm max})$ overlap within the
error bars.  
%%%
(b) $|\alpha_P|$ as a
  function 
%%%
of 
%%%
$T$ for different $P$. 
(c) $|dp_{\rm
  B}/dT|$, the temperature derivative of $p_{\rm B}$, as a function of
  $T$ for different $P$. In panel (a), 
%%%
$T(|dp_{\rm B}/dT|^{\rm max})$
%%%
  ($\Diamond$) overlaps $T_W(P)$. 
}

\label{locus}
\end{figure}

\begin{figure}
\begin{center}
\includegraphics[width=8cm]{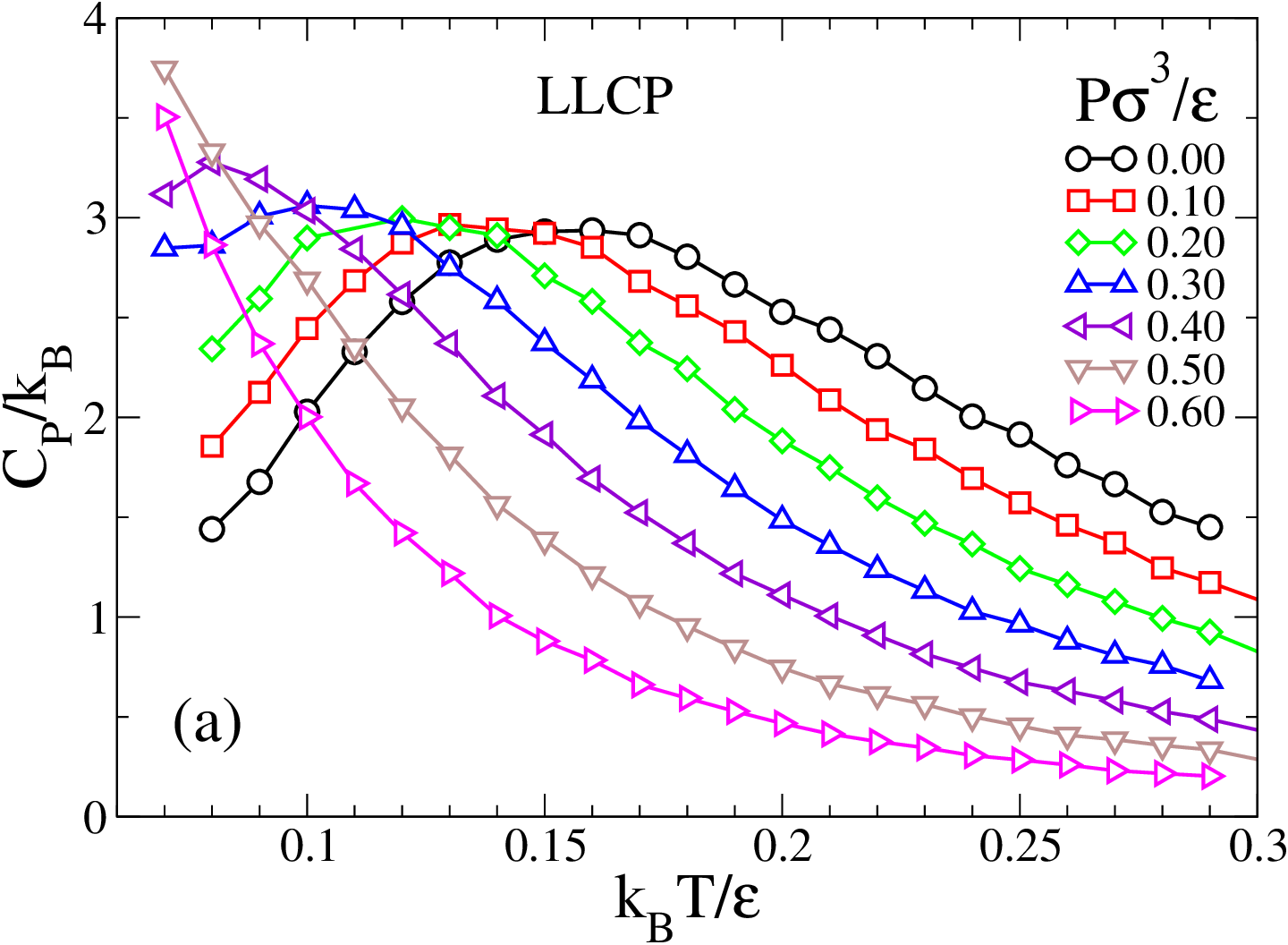}
\includegraphics[width=8cm]{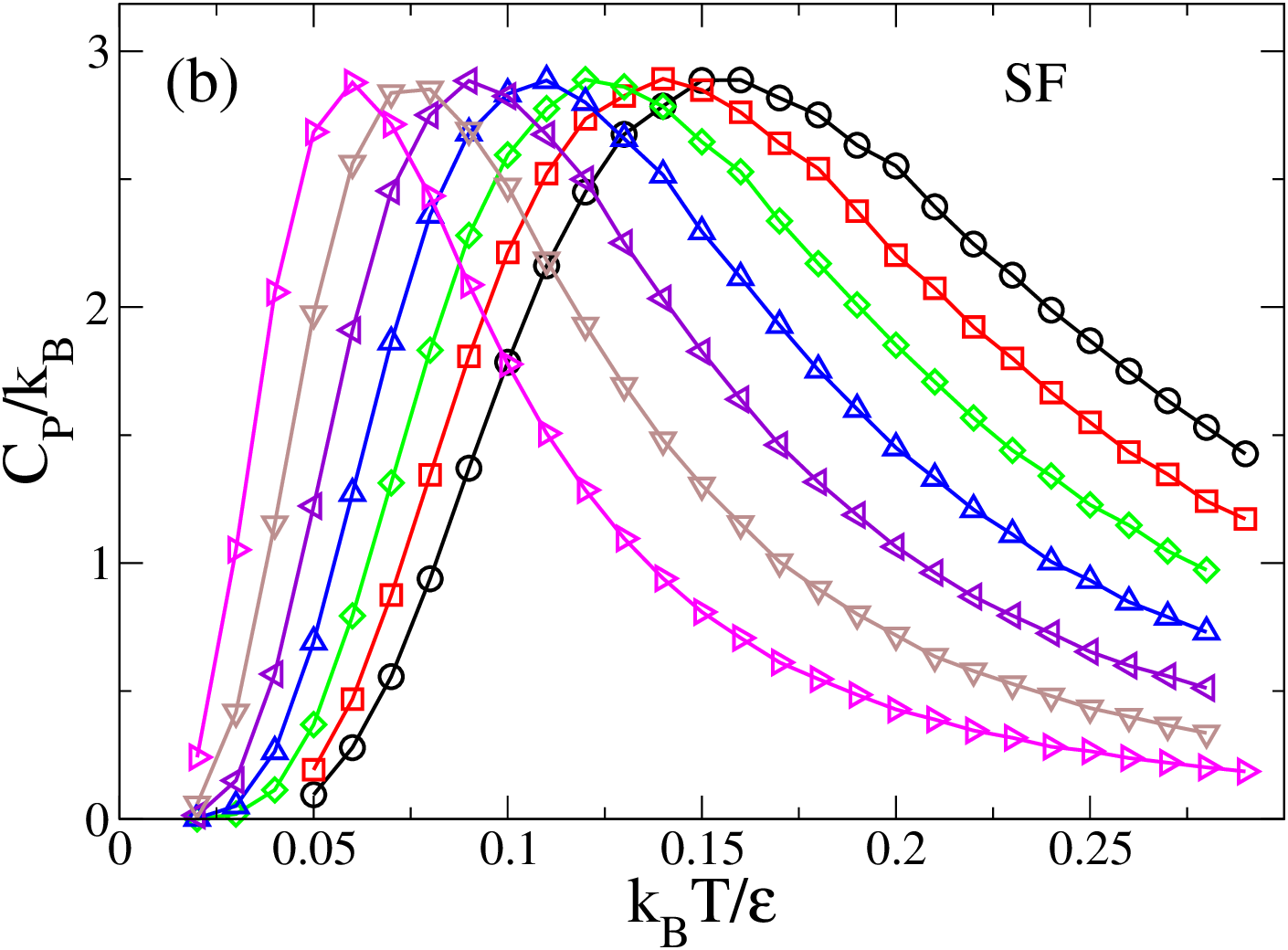}
\end{center}
\caption{ (color online) Temperature dependence of the specific heat
  $C_P$ for both the LLCP and the SF scenarios, for seven values of
  $P$. (a) For the LLCP scenario, $C_P$ has a maximum, the size of which
  increases with increasing pressure and diverges as $P\rightarrow
  P_{C'}$. (b) For the SF scenario, $C_P$ also has a maximum, but its
  size does not increase with increasing pressure, consistent with the
  findings of the mean field calculations of Ref.~\cite{sdss}.}
\label{Cp}
\end{figure}

Two different scenarios are commonly used to interpret the anomalies of
 water~\cite{angellreview,pabloreview}: 

\begin{itemize}

\item The {\it liquid-liquid critical point (LLCP)} scenario
      hypothesizes that supercooled water has a liquid-liquid phase
      transition line that separates a low-density liquid (LDL) at low
      temperature $T$ and low pressure $P$ and a high-density liquid
      (HDL) at high $T$ and $P$ and terminates at a critical point $C'$
      \cite{Mishima1998nature}.  From $C'$ emanates the Widom line
      $T_W(P)$, the line of maximum correlation length in the $(T,P)$
      plane.
Response
      functions, such as the isobaric heat capacity $C_P$, the
      coefficient of thermal expansion $\alpha_P$, and the isothermal
      compressibility $K_T$, have maxima along lines that converge
      toward $T_W(P)$ upon approaching $C'$
      [Figs.~\ref{locus} and \ref{Cp}(a)].

\item The {\it singularity-free (SF) } scenario hypothesizes the
      presence of a line of temperatures of maximum density $T_{\rm MD}(P)$
      with negative slope in the $(T,P)$ plane.  As a consequence, $K_T$
      and $|\alpha_P|$ have maxima that increase upon increasing $P$, as
      shown using a cell model of water. The maxima in $C_P$ do
      not increase with $P$, suggesting that there is no singularity
      \cite{sdss} [Fig.~\ref{Cp}(b)].

\end{itemize}

Above the homogeneous nucleation line $T_H(P)$ where data are available,
the two scenarios predict roughly the same equilibrium phase diagram.
  Here we show that {\it dynamic} measurements should reveal a striking
  difference between the two scenarios.  Specifically, the low-$T$
  dynamics depends on local structural changes, quantified by the
  variation of the number of hydrogen bonds, that are affected by
  pressure differently for each scenario.  We find this result by
  studying---using Monte Carlo (MC) simulations and mean field
  calculations---a cell model which has the property that by tuning a
  parameter its predictions conform to those of either the LLCP or the
  SF scenario.  This cell model is based on the experimental
  observations that on decreasing $P$ at constant $T$, or on decreasing
  $T$ at constant $P$, (i) water displays an increasing local
  tetrahedrality \cite{Darrigo81}, (ii) the volume per molecule
  increases at sufficiently low $P$ or $T$, and (iii) the O-O-O angular
  correlation increases \cite{Bosio83}.

The entire system is divided into cells $i\in[1,\ldots,N]$, each containing a molecule with volume $v\equiv V/N$, where
$V\geq N v_{\rm hc}$ is the total volume of the system, and $v_{\rm hc}$ is the
hard-core volume of one molecule. The cell volume $v$ is a continuous
variable that gives, in $d$ dimensions, the mean distance $r\equiv
v^{1/d}$ between molecules.  The van der Waals interaction is represented
by a potential with attractive energy $\epsilon>0$ between
nearest-neighbor (n.n.) molecules and a hard-core repulsion at
%%%
$R_{\rm hc}\equiv v_{\rm hc}^{1/d}$.
%%%

For a regular square lattice, each molecule $i$ has four bond indices
$\sigma_{ij} \in [1,\ldots,q]$, corresponding to the four n.n. cells
$j$, giving rise to $q^4$ different molecular orientations.  Bonding and
intramolecular (IM) interactions are accounted for by the two
Hamiltonian terms
\begin{equation}
{\cal H}_{\rm B}
\equiv -J\sum_{\langle i,j\rangle}
\delta_{\sigma_{ij}\sigma_{ji}} ,
\label{HB}
\end{equation}
where the sum is over n.n. cells, $0<J<\epsilon$ is the bond
energy, $\delta_{a,b}=1$ if $a=b$ and $\delta_{a,b}=0$ otherwise, and 
\begin{equation}
{\cal H}_{\rm IM}\equiv -J_{\sigma} \sum_{i} \sum_{(k,\ell)_i}
\delta_{\sigma_{ik}\sigma_{i\ell}},
\label{HIM}
\end{equation}
where $\sum_{(k,\ell)_i}$ denotes the sum over the IM bond indices
$(k,l)$ of the molecule $i$ and $J_\sigma>0$ is the IM interaction
energy with $J_{\sigma}<J$, which models the angular correlation between
the bonds on the same molecule. The total energy of the system is the
sum of the van der Waals interaction and Eqs.~(\ref{HB}) and (\ref{HIM}).

At constant $P$, the density of water decreases for $T<T_{\rm MD}(P)$
which the model takes into account by increasing the total volume by an
amount $v_{\rm B}>0$ for each bond formed.  Hence the total molar volume
$v$ of the system is
\begin{equation}
v=v_{\rm free}+ 2p_{\rm B} v_{\rm B},
\label{eq:v}
\end{equation}
where $v_{\rm free}$ is 
%%%
a variable for
%%%
the molar volume without taking into account the
bonds, $p_{\rm B}=N_B/(2N)$ is the fraction of bonds formed and $N_B$ is
the number of bonds \cite{fs,sdss}.

We perform simulations in the $NPT$ ensemble \cite{fs} for $q=6$,
$v_{\rm B}/v_{\rm hc}=0.5$, $J/\epsilon=0.5$, and for two different values of
$J_{\sigma}/\epsilon$: (i) $J_\sigma/\epsilon=0.05$, which gives rise to
a phase diagram with a LLCP [Fig.~\ref{locus}(a)], and (ii)
$J_\sigma=0$, which leads to the SF scenario \cite{sdss}. We study two
square lattices with 900 and 3600 cells, and find no appreciable size
effects. We collect statistics over $10^6$ MC steps after equilibrating
the system for all $P$ and $T$.

For $J_\sigma/\epsilon=0.05$, $|\alpha_P|$ for $P<P_{C'}$ displays a
maximum, $\alpha_P^{\rm max}$ [Fig.~\ref{locus}(b)].  As $P$ 
increases, $\alpha_P^{\rm max}$ increases and shifts to lower $T$,
converging toward $T_W(P)$ [Fig.~\ref{locus}(a)]. We find that the
number of bonds, $N_B$, increases on decreasing $T$, and at constant $T$
decreases for increasing $P$, and is almost constant at $T_W(P)$
\cite{0.8S}. This is consistent with trends seen both in experiments
\cite{Darrigo81} and in simulations
\cite{Kumar07PNAS}, suggesting that for
$T>T_W(P)$ the liquid is less structured and more HDL-like, while for
$T<T_W(P)$ it is more structured and more LDL-like.

 We find that $|dp_{\rm B}/dT|$ shows a clear maximum for all $P<P_{C'}$
which shifts to lower $T$ upon increasing $P$ [Fig.~\ref{locus}(c)].
Remarkably, we also find that the locus of $|dp_{\rm B}/dT|^{\rm max}$
coincides with the Widom line $T_W(P)$ [Fig.~\ref{locus}(a)] and that
the value of $|dp_{\rm B}/dT|^{\rm max}$ increases on approaching
$P_{C'}$. This is the same qualitative behavior as $|\alpha_P(T)|^{\rm
max}$ and $C_P(T)^{\rm max}$, which are used to locate $T_W(P)$
[Figs.~\ref{locus}(b) and \ref{Cp}(a)].  The relation of $|dp_{\rm
B}/dT|$ with the fluctuations is revealed by its proportionality to
$|\alpha_P(T)|$ and to the fluctuation of the number of bonds
\begin{equation}
\langle N_B^2\rangle-\langle
N_B\rangle^2=\frac{2Nk_BT^2}{J-Pv_B}\left|\frac{dp_{\rm B}}{dT}\right| ,
\end{equation}
where $k_B$ is the Boltzmann constant.

For $J_\sigma=0$ (SF scenario) we observe no difference for the behavior
of $N_B$ and $|dp_{\rm B}/dT|$. We further verify the prediction of the
SF scenario \cite{sdss} that $C_P^{\rm max}$ remains a constant upon
increasing $P$ [Fig.~\ref{Cp}(b)].

Next, we study how this different behavior affects the dynamics.
Previous simulations \cite{Xu2005} found a crossover from
non-Arrhenius to Arrhenius dynamics for the diffusion constant of models
that display a LLCP, and showed the temperature of this crossover
coincided with $T_W(P)$.  We calculate, for both scenarios, the
relaxation time $\tau$ of $S_i\equiv\sum_j\sigma_{ij}/4$, which
quantifies the degree of total bond ordering for site $i$.
Specifically, we identify $\tau$ as the time for the spin
autocorrelation function $C_{\sigma \sigma} (t)\equiv\langle
S_i(t)S_i(0)\rangle/\langle S_i^2(0)\rangle$ to decay to the value $1/e$.

\begin{figure}
\begin{center}
\includegraphics[width=8cm]{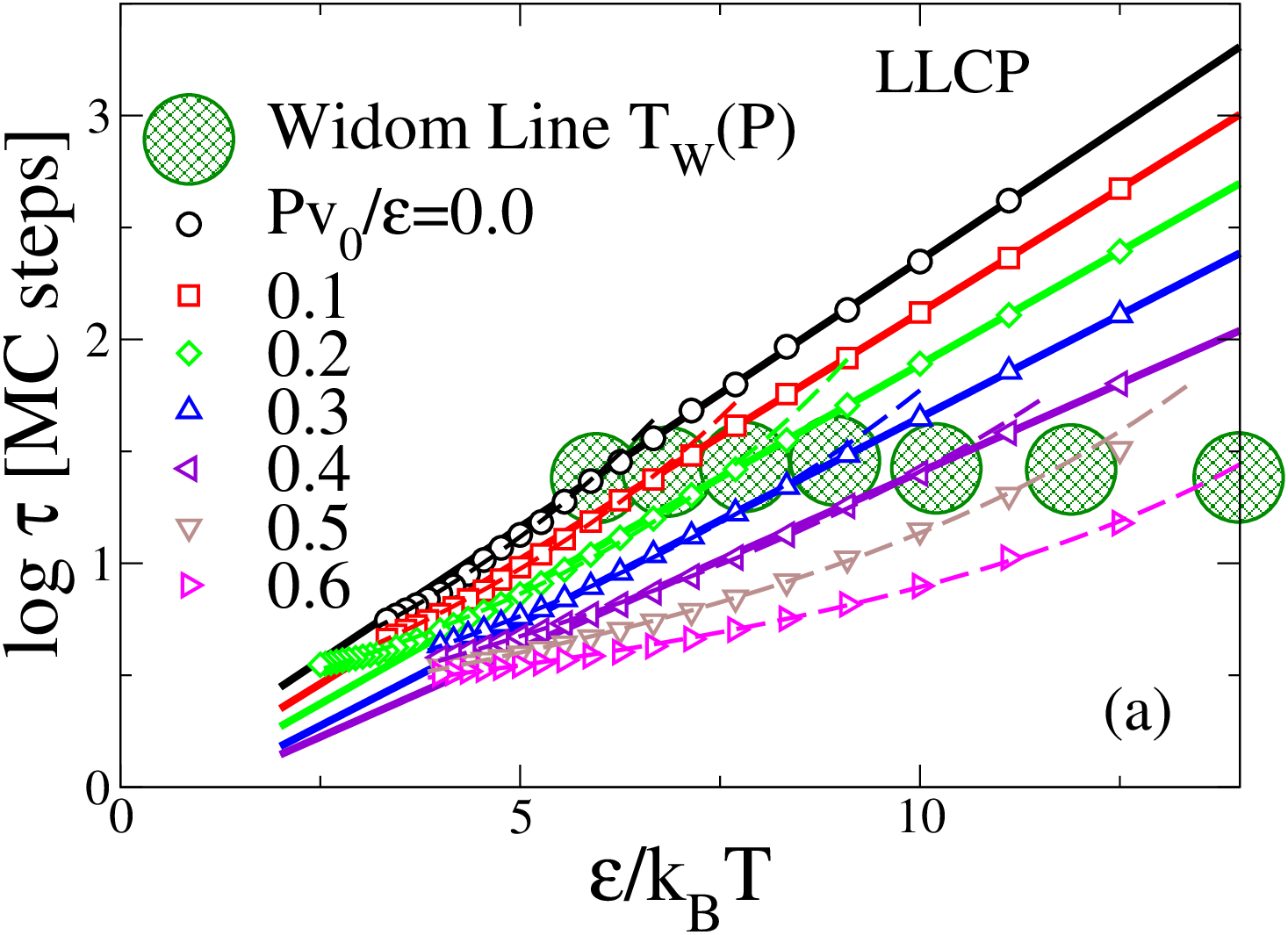}
\includegraphics[width=8cm]{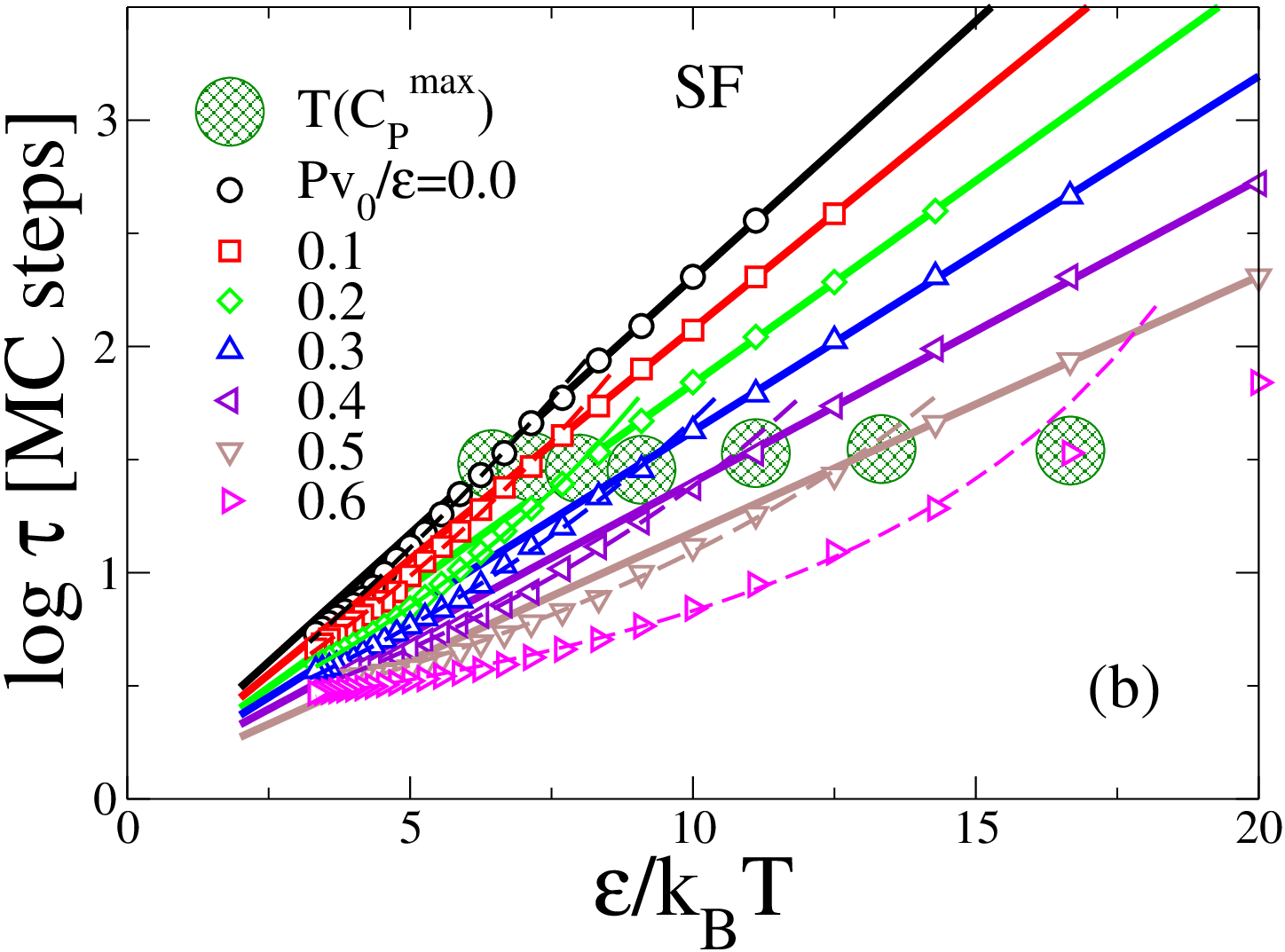}
\end{center}
\caption{Dynamic crossover---large hatched circles of a radius
 approximately equal to the error bar---in the orientational relaxation
 time $\tau$ for a range of different pressures. (a) The LLCP scenario,
 with crossover temperature at $T_W(P)$.  (b) The SF scenario, with
 crossover temperature at $T(C_P^{\rm max})$.   Solid and dashed
 lines represent Arrhenius and VFT fits, respectively.  Notice that 
 the dynamic crossover occurs at approximately the same value of $\tau$
 for all seven values of pressure studied.
}
\label{tau}
\end{figure}

For both scenarios we find a dynamic crossover  (Fig.~\ref{tau}).  At
high $T$, we fit $\tau$ with the Vogel-Fulcher-Tamman (VFT) function
\begin{equation}
\tau^{\rm VFT}\equiv\tau_0^{\rm VFT} \exp\left[{T_1\over T-T_0}\right],
\label{VFT}
\end{equation}
where $\tau_0^{\rm VFT}$, $T_1$, and $T_0$ are three fitting parameters.
We find that $\tau$ has an Arrhenius $T$ dependence at low $T$,
$\tau=\tau_0\exp[E_{\rm A}/k_BT]$, where $\tau_0$ is the relaxation time
in the high-$T$ limit, and $E_{\rm A}$ is a $T$-independent activation
energy. We find that for $J_\sigma/\epsilon=0.05$ the crossover occurs
at $T_W(P)$ for $P<P_{C'}$ [Fig.~\ref{tau}(a)], and that for
$J_\sigma=0$ the crossover is at $T(C_P^{\rm max})$, the temperature of
$C_P^{\rm max}$ [Fig.~\ref{tau}(b)].  We note that for both scenarios
the crossover is isochronic, 
%%%
i.e. the value of the crossover time $\tau_{\rm C}$ is
approximately independent of pressure.
We find $\tau_{\rm C}\simeq 10^{3/2}$ MC steps.
%%%  

\begin{figure}
\begin{center}
\includegraphics[width=8cm]{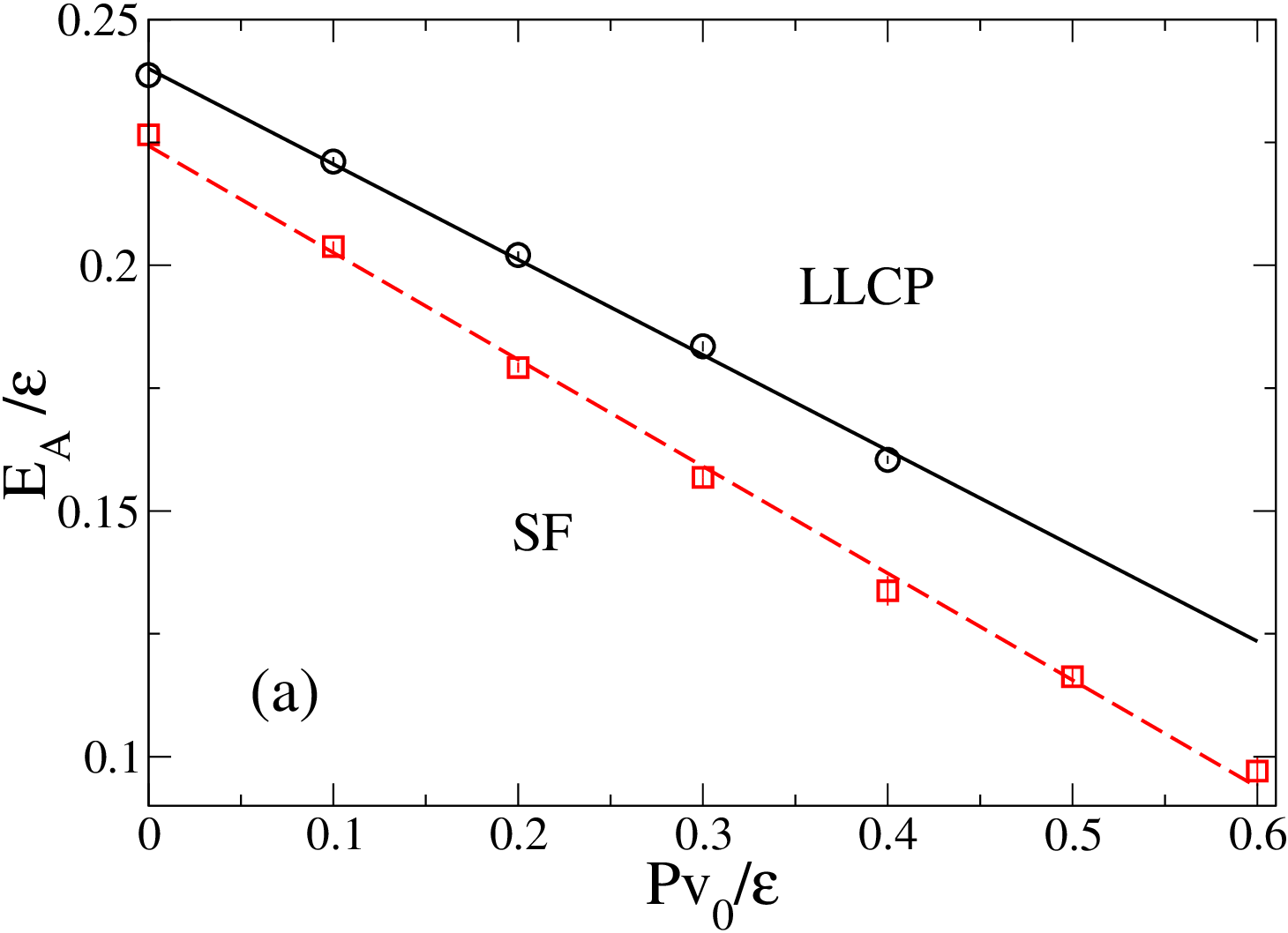}
\includegraphics[width=8cm]{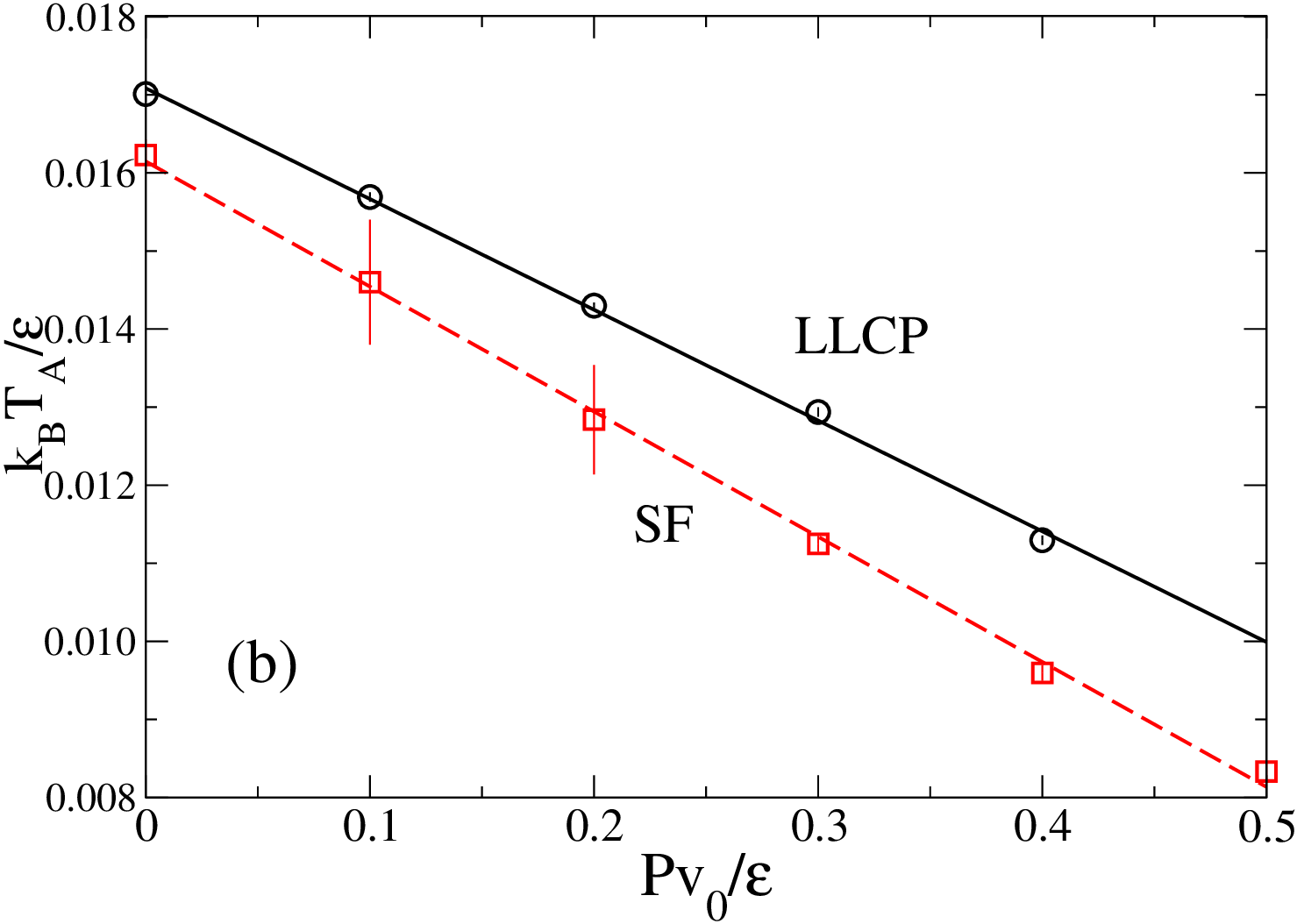}
\includegraphics[width=8cm]{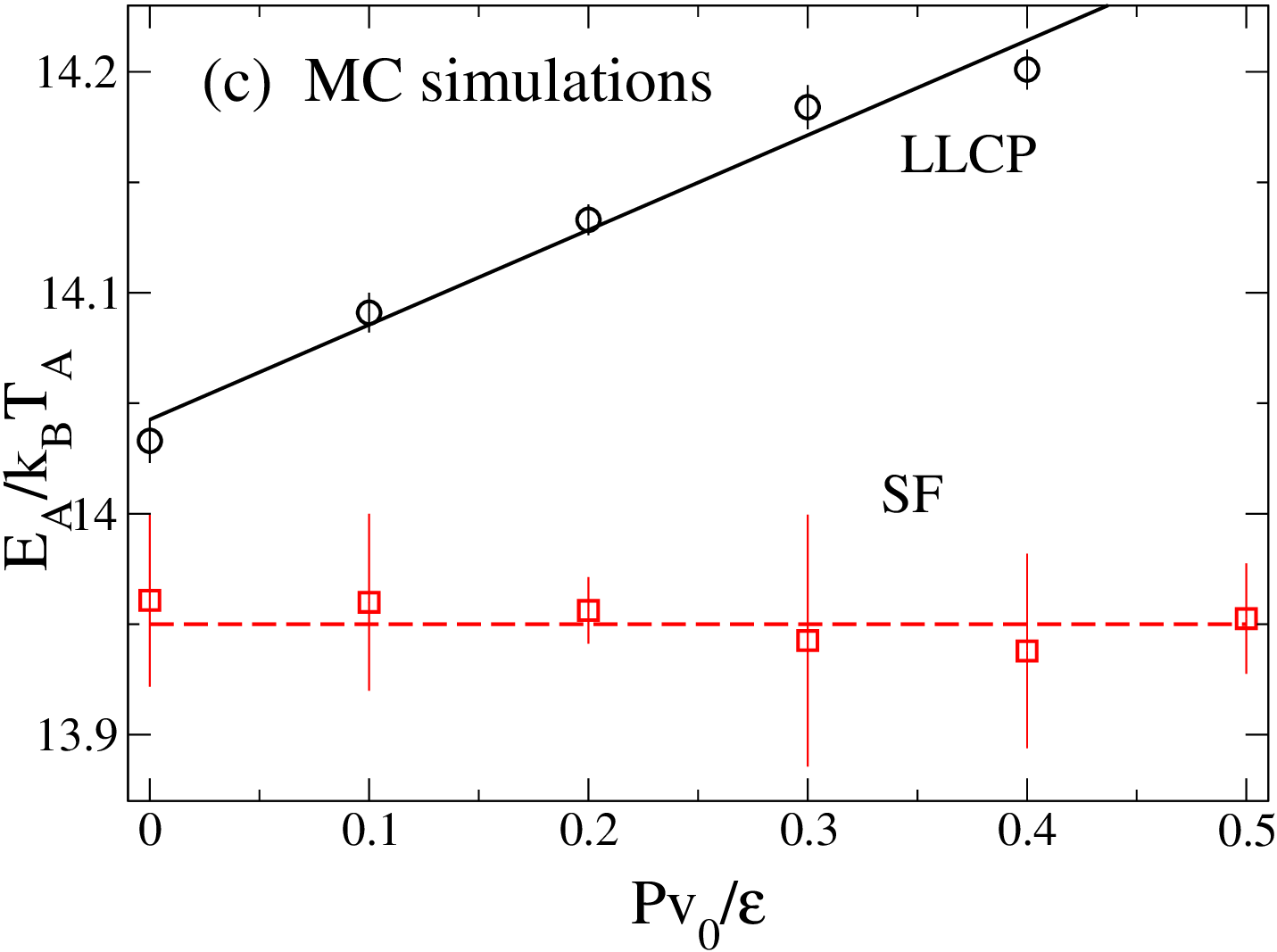}
\includegraphics[width=8cm]{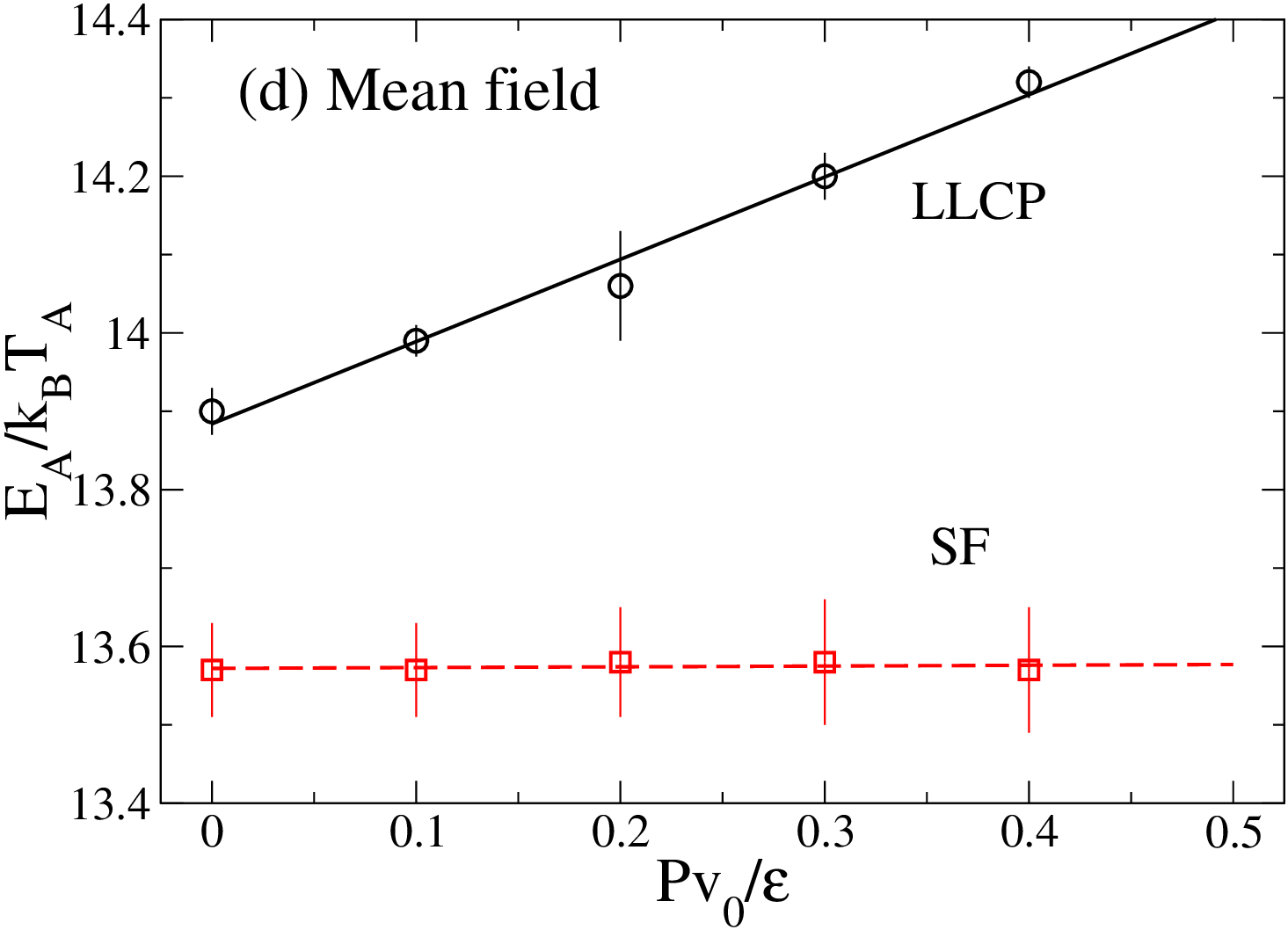}
\end{center}
\caption{Effect of pressure on the activation energy $E_{\rm A}$.  (a)
  Demonstration that $E_{\rm A}$ decreases linearly for increasing $P$
  for both the LLCP and the SF scenarios. The lines are linear fits to
  the simulation results (symbols). (b) $T_{\rm A}$, defined such that
  $\tau(T_{\rm A})=10^{14}$ MC steps $> 100$~sec \cite{kfbs06},
  decreases linearly with $P$ for both scenarios. (c) $P$ dependence of
  the quantity $E_{\rm A}/(k_BT_{\rm A})$ is different in the two
  scenarios.  In the LLCP scenario, $E_{\rm A}/(k_BT_{\rm A})$ increases
  with increasing $P$, and it is approximately constant in the SF scenario.
  The lines are guides to the eyes.  (d) Demonstration that the same
  behavior is found using the mean field approximation.  In all the
  panels, where not shown, the error bars are smaller than the symbol
  sizes.  }
\label{Tg}
\end{figure}

We next calculate the Arrhenius activation energy $E_{\rm A}(P)$ from
the low-$T$ slope of $\log \tau$ vs. $1/T$ [Fig.~\ref{Tg}(a)].  We
extrapolate the temperature $T_{\rm A}(P)$ at which $\tau$ reaches a
fixed macroscopic time $\tau_{\rm A}\geq \tau_{\rm C}$.
%, with $T_{\rm A}(P)$ smaller than the crossover temperature.  
We choose
$\tau_{\rm A}=10^{14}$ MC steps $> 100$~sec \cite{kfbs06}
[Fig.~\ref{Tg}(b)].  We find that $E_{\rm A}(P)$ and $T_{\rm A}(P)$ decrease
upon increasing $P$ in both scenarios, providing no distinction between
the two interpretations. Instead, we find a dramatic difference in the
$P$ dependence of the quantity $E_{\rm A}/(k_B T_{\rm A})$ in the two
scenarios, increasing for the LLCP scenario and approximately constant for
the SF scenario [Fig.~\ref{Tg}(c)].

We can better understand our findings by developing an expression for
$\tau$ in terms of thermodynamic quantities, which will then allow us to
explicitly calculate $E_{\rm A}/(k_B T_{\rm A})$ for both scenarios.  For
any activated process, in which the relaxation from an initial state to
a final state passes  through an excited transition state,
%$\tau=\tau_0\exp[\Delta(U+PV-TS)/k_BT]$, 
$\ln (\tau/\tau_0)=\Delta(U+PV-TS)/(k_BT)$, 
where $\Delta(U+PV-TS)$ is the
difference in free energy between the transition state and the initial
state.  Consistent with results from simulations and experiments
\cite{Laage-Hynes2006,Tokmakoff}, we propose that at low $T$ the
mechanism to relax from a less structured state (lower tetrahedral
order) to a more structured state (higher tetrahedral order) corresponds
to the breaking of a bond and the simultaneous molecular reorientation
for the formation of a new bond.  The transition state is represented by
the molecule with a broken bond and more  tetrahedral IM order.  Hence,
\begin{equation}
\Delta(U+PV-TS) 
=Jp_{\rm B}-J_{\sigma} p_{\rm IM} - Pv_{\rm B} - T \Delta S, 
\label{DG}
\end{equation}
where $p_{\rm B}$ and $p_{\rm IM}$, the probability of a satisfied IM
interaction, can be directly calculated.  To estimate $\Delta S$, the
increase of entropy due to the breaking of a bond, we use the mean field
expression $\Delta S=k_B[\ln(2Np_{\rm B})-\ln(1+2N(1-p_{\rm
B}))]\bar{p}_{\rm B}$, where $\bar{p}_{\rm B}$ is the average value of
$p_{\rm B}$ above and below $T_W(P)$.

We next test 
%%%
that the expression of 
$\ln (\tau/\tau_0)$, in terms of $\Delta S$ and Eq.(\ref{DG}),
%%%
%
\begin{equation}
\ln \frac{\tau}{\tau_0}= 
\frac{
Jp_{\rm B} - J_{\sigma}p_{\rm IM}  - Pv_{\rm B}
}{k_BT} - 
\bar{p}_{\rm B}\ln \frac{2Np_{\rm B}}{1+2N(1-p_{\rm B})}
\label{relation}
\end{equation}
describes the simulations well, with minor corrections at high $T$.
Here  $\tau_0 \equiv\tau_0(P)$ is a free fitting parameter equal to the
 relaxation time for $T\to \infty$.  From Eq.(\ref{relation}) we
find that the ratio $E_{\rm A}/(k_B T_{\rm A})$ calculated at low $T$
increases with $P$ for $J_\sigma/\epsilon=0.05$, while it is constant
for $J_\sigma=0$, as from our simulations [Fig.~\ref{Tg}(d)]. 
 
In summary, we have seen that both the LLCP and SF scenarios exhibit a
dynamic crossover at a temperature close to $T(C_P^{\rm max})$, which
decreases for increasing $P$.  We interpret the dynamic crossover as a
consequence of a local breaking and reorientation of the bonds for the
formation of new and more tetrahedrally oriented bonds.  Above
$T(C_P^{\rm max})$, when $T$ decreases, the number of hydrogen bonds
increases, giving rise to an increasing activation energy $E_{\rm A}$
and to a non-Arrhenius dynamics.  As $T$ decreases, entropy must
decrease.  A major contributor to entropy is the orientational disorder,
%so it is plausible that 
%%%
that is a function of $p_{\rm B}$, as described by the mean field
expression for $\Delta S$.
We find that, as $T$ decreases, $p_{\rm B}$ --- hence the
orientational order ---
increases.  We find that the rate of increase has a maximum at
$T(C_P^{\rm max})$, 
and as $T$ continues to decrease this rate drops rapidly to zero ---
meaning that for $T<T(C_P^{\rm max})$, the local
orientational order rapidly becomes temperature-independent and the
activation energy $E_{\rm A}$ 
%plausibly 
also becomes approximately
temperature-independent, for the Eq.(\ref{DG}).  Corresponding
to this fact the dynamics 
becomes approximately Arrhenius.
%%%

We find that the crossover is approximately isochronic (independent of
the pressure) consistent with our calculations of an almost constant
number of bonds at $T(C_P^{\rm max})$. In both scenarios, $E_{\rm A}$
and $T_{\rm A}$ decrease upon increasing $P$, but the $P$ dependence of
the quantity $E_{\rm A}/(k_B T_{\rm A})$ has a dramatically different
behavior in the two scenarios. For the LLCP scenario it increases as
$P\to P_{C'}$, while it is approximately constant in the SF scenario.
%%%
We interpret this difference as a consequence of the larger increase of
the rate of change of $p_{\rm B}$ in the LLCP scenario, where $p_{\rm
B}$ diverges at finite $T_{C'}$, compared to the SF scenario, where
$p_{\rm B}$ can possibly diverge only at $T=0$.   Since experiments can
detect local changes of water structure from HDL-like to LDL-like,
(e. g. \cite{Li05}), it is possible that our prediction on the dynamic
consequences of this local change may be experimentally testable.
%%%

%This prediction can be tested experimentally.  For example, recent
%experiments \cite{Li05} interpreted the variation of the sound velocity
%pressure response at moderate temperatures (between 293~K and 353~K) and
%high $P$ (between 0.19~GPa and 0.29~GPa) as the result of a
%change of
%structure of water , consistent with numerical simulations
%showing the 

%continuous but sharp variation from HDL-like local structure of water to
%LDL-likelocal structure \cite{Saitta03}.

\bigskip

\noindent We thank C. A. Angell, M.-C. Bellissent-Funel, W. Kob, L. Liu
and S. Sastry for helpful discussions and NSF grant CHE0616489 for
support. G. F. also thanks the Spanish Ministerio de Educaci\'on y
Ciencia (Programa Ram\'on y Cajal and Grant No. FIS2004-03454).

%\pagebreak
%\begin{widetext}

\begin{thebibliography}{99}
\bibitem{0.8S} We find $p_B\simeq 0.8$ at $T_W(P)$, being somewhat
  smaller for increasing $P$, consistent with the value 0.795 used in
  \cite{sdss} to trace the temperature of maximum $K_T$ for $J_\sigma=
  0$. We find that for $J_\sigma/\epsilon=0.05$ polynomial
  extrapolations of $p_{\rm B}$ at $C'$ up to the fifth order lead to
  a value of $p_{\rm B}(C')=0.55\pm 0.15$. Mean field calculations of
  $p_{\rm B}$ compare well with simulations for $T\gtrsim T_W(P)$ and
  for $T\ll T_W(P)$.
\end{thebibliography}
\end{document}